 \input pictex.tex   
\immediate\write10{Package DCpic 2002/05/16 v4.0}

\catcode`!=11 

\newcount\aux%
\newcount\auxa%
\newcount\auxb%
\newcount\m%
\newcount\n%
\newcount\x%
\newcount\y%
\newcount\xl%
\newcount\yl%
\newcount\d%
\newcount\dnm%
\newcount\xa%
\newcount\xb%
\newcount\xmed%
\newcount\xc%
\newcount\xd%
\newcount\ya%
\newcount\yb%
\newcount\ymed%
\newcount\yc%
\newcount\yd
\newcount\expansao%
\newcount\tipografo
\newcount\distanciaobjmor
\newcount\tipoarco
\newif\ifpara%
\newbox\caixa%
\newbox\caixaaux%
\newif\ifnvazia%
\newif\ifvazia%
\newif\ifcompara%
\newif\ifdiferentes%
\newcount\xaux%
\newcount\yaux%
\newcount\guardaauxa%
\newcount\alt%
\newcount\larg%
\newcount\prof%
\newcount\auxqx
\newcount\auxqy
\newif\ifajusta%
\newif\ifajustadist
\def\objPartida{}%
\def\objChegada{}%
\def\objNulo{}%


\def\!vazia{:}

\def\!pilhanvazia#1{\let\arg=#1%
\if:\arg\ \nvaziafalse\vaziatrue \else \nvaziatrue\vaziafalse\fi}

\def\!coloca#1#2{\edef\pilha{#1.#2}}

\def\!guarda(#1)(#2,#3)(#4,#5,#6){\def\id{#1}%
\xaux=#2%
\yaux=#3%
\alt=#4%
\larg=#5%
\prof=#6%
}

\def\!topaux#1.#2:{\!guarda#1}
\def\!topo#1{\expandafter\!topaux#1}

\def\!popaux#1.#2:{\def\pilha{#2:}}
\def\!retira#1{\expandafter\!popaux#1}

\def\!comparaaux#1#2{\let\argA=#1\let\argB=#2%
\ifx\argA\argB\comparatrue\diferentesfalse\else\comparafalse\diferentestrue\fi}

\def\!compara#1#2{\!comparaaux{#1}{#2}}

\def\!absoluto#1#2{\n=#1%
  \ifnum \n > 0
    #2=\n
  \else
    \multiply \n by -1
    #2=\n
  \fi}


\def\dasharrow{1}
\def\solidline{2}
\def\injectionarrow{3}


\def\atleft{1}

\def\commdiag{0}


\def\!ajusta#1#2#3#4#5#6{\aux=#5%
  \let\auxobj=#6%
  \ifcase \tipografo    
    \ifnum\number\aux=10 
      \ajustadisttrue 
    \else
      \ajustadistfalse  
    \fi
  \else  
   \ajustadistfalse
  \fi
  \ifajustadist
   \let\pilhaaux=\pilha%
   \loop%
     \!topo{\pilha}%
     \!retira{\pilha}%
     \!compara{\id}{\auxobj}%
     \ifcompara\nvaziafalse \else\!pilhanvazia\pilha \fi%
     \ifnvazia%
   \repeat%
   \let\pilha=\pilhaaux%
   \ifvazia%
    \ifdiferentes%
     \larg=1310720
     \prof=655360%
     \alt=655360%
    \fi%
   \fi%
   \divide\larg by 131072
   \divide\prof by 65536
   \divide\alt by 65536
   \ifnum\number\y=\number\yl
    \advance\larg by 3
    \ifnum\number\larg>\aux
     #5=\larg
    \fi
   \else
    \ifnum\number\x=\number\xl
     \ifnum\number\yl>\number\y
      \ifnum\number\alt>\aux
       #5=\alt
      \fi
     \else
      \advance\prof by 5
      \ifnum\number\prof>\aux
       #5=\prof
      \fi
     \fi
    \else
     \auxqx=\x
     \advance\auxqx by -\xl
     \!absoluto{\auxqx}{\auxqx}%
     \auxqy=\y
     \advance\auxqy by -\yl
     \!absoluto{\auxqy}{\auxqy}%
     \ifnum\auxqx>\auxqy
      \ifnum\larg<10
       \larg=10
      \fi
      \advance\larg by 3
      #5=\larg
     \else
      \ifnum\yl>\y
       \ifnum\larg<10
        \larg=10
       \fi
      \advance\alt by 6
       #5=\alt
      \else
      \advance\prof by 11
       #5=\prof
      \fi
     \fi
    \fi
   \fi
\fi} 

\def\!raiz#1#2{\n=#1%
  \m=1%
  \loop
    \aux=\m%
    \advance \aux by 1%
    \multiply \aux by \aux%
    \ifnum \aux < \n%
      \advance \m by 1%
      \paratrue%
    \else\ifnum \aux=\n%
      \advance \m by 1%
      \paratrue%
       \else\parafalse%
       \fi
    \fi
  \ifpara%
  \repeat
#2=\m}

\def\!ucoord#1#2#3#4#5#6#7{\aux=#2%
  \advance \aux by -#1%
  \multiply \aux by #4%
  \divide \aux by #5%
  \ifnum #7 = -1 \multiply \aux by -1 \fi%
  \advance \aux by #3%
#6=\aux}

\def\!quadrado#1#2#3{\aux=#1%
  \advance \aux by -#2%
  \multiply \aux by \aux%
#3=\aux}

\def\!distnomemor#1#2#3#4#5#6{\setbox0=\hbox{#5}%
  \aux=#1
  \advance \aux by -#3
  \ifnum \aux=0
     \aux=\wd0 \divide \aux by 131072
     \advance \aux by 3
     #6=\aux
  \else
     \aux=#2
     \advance \aux by -#4
     \ifnum \aux=0
        \aux=\ht0 \advance \aux by \dp0 \divide \aux by 131072
        \advance \aux by 3
        #6=\aux%
     \else
     #6=3
     \fi
   \fi
}

\def\begindc#1{\!ifnextchar[{\!begindc{#1}}{\!begindc{#1}[30]}}
\def\!begindc#1[#2]{\beginpicture 
  \let\pilha=\!vazia
  \setcoordinatesystem units <1pt,1pt>
  \expansao=#2
  \ifcase #1
    \distanciaobjmor=10
    \tipoarco=0         
    \tipografo=0        
  \or
    \distanciaobjmor=2
    \tipoarco=0         
    \tipografo=1        
  \or
    \distanciaobjmor=1
    \tipoarco=2         
    \tipografo=2        
  \or
    \distanciaobjmor=8
    \tipoarco=0         
    \tipografo=3        
  \or
    \distanciaobjmor=8
    \tipoarco=2         
    \tipografo=4        
  \fi}

\def\enddc{\endpicture}

\def\mor{%
  \!ifnextchar({\!morxy}{\!morObjA}}
\def\!morxy(#1,#2){%
  \!ifnextchar({\!morxyl{#1}{#2}}{\!morObjB{#1}{#2}}}
\def\!morxyl#1#2(#3,#4){%
  \!ifnextchar[{\!mora{#1}{#2}{#3}{#4}}{\!mora{#1}{#2}{#3}{#4}[\number\distanciaobjmor,\number\distanciaobjmor]}}%
\def\!morObjA#1{%
 \let\pilhaaux=\pilha%
 \def\objPartida{#1}%
 \loop%
    \!topo\pilha%
    \!retira\pilha%
    \!compara{\id}{\objPartida}%
    \ifcompara \nvaziafalse \else \!pilhanvazia\pilha \fi%
   \ifnvazia%
 \repeat%
 \ifvazia%
  \ifdiferentes%
   Error: Incorrect label specification%
   \xaux=1%
   \yaux=1%
  \fi%
 \fi%
 \let\pilha=\pilhaaux%
 \!ifnextchar({\!morxyl{\number\xaux}{\number\yaux}}{\!morObjB{\number\xaux}{\number\yaux}}}
\def\!morObjB#1#2#3{%
  \x=#1
  \y=#2
 \def\objChegada{#3}%
 \let\pilhaaux=\pilha%
 \loop
    \!topo\pilha %
    \!retira\pilha%
    \!compara{\id}{\objChegada}%
    \ifcompara \nvaziafalse \else \!pilhanvazia\pilha \fi
   \ifnvazia
 \repeat
 \ifvazia
  \ifdiferentes%
   Error: Incorrect label specification
   \xaux=\x%
   \advance\xaux by \x%
   \yaux=\y%
   \advance\yaux by \y%
  \fi
 \fi
 \let\pilha=\pilhaaux
 \!ifnextchar[{\!mora{\number\x}{\number\y}{\number\xaux}{\number\yaux}}{\!mora{\number\x}{\number\y}{\number\xaux}{\number\yaux}[\number\distanciaobjmor,\number\distanciaobjmor]}}
\def\!mora#1#2#3#4[#5,#6]#7{%
  \!ifnextchar[{\!morb{#1}{#2}{#3}{#4}{#5}{#6}{#7}}{\!morb{#1}{#2}{#3}{#4}{#5}{#6}{#7}[1,\number\tipoarco] }}
\def\!morb#1#2#3#4#5#6#7[#8,#9]{\x=#1%
  \y=#2%
  \xl=#3%
  \yl=#4%
  \multiply \x by \expansao%
  \multiply \y by \expansao%
  \multiply \xl by \expansao%
  \multiply \yl by \expansao%
  \!quadrado{\number\x}{\number\xl}{\auxa}%
  \!quadrado{\number\y}{\number\yl}{\auxb}%
  \d=\auxa%
  \advance \d by \auxb%
  \!raiz{\d}{\d}%
  \auxa=#5
  \!compara{\objNulo}{\objPartida}%
  \ifdiferentes
   \!ajusta{\x}{\xl}{\y}{\yl}{\auxa}{\objPartida}%
   \ajustatrue
   \def\objPartida{}
  \fi
  \guardaauxa=\auxa
  \!ucoord{\number\x}{\number\xl}{\number\x}{\auxa}{\number\d}{\xa}{1}%
  \!ucoord{\number\y}{\number\yl}{\number\y}{\auxa}{\number\d}{\ya}{1}%
  \auxa=\d%
  \auxb=#6
  \!compara{\objNulo}{\objChegada}%
  \ifdiferentes
   \!ajusta{\x}{\xl}{\y}{\yl}{\auxb}{\objChegada}%
   \def\objChegada{}
  \fi
  \advance \auxa by -\auxb%
  \!ucoord{\number\x}{\number\xl}{\number\x}{\number\auxa}{\number\d}{\xb}{1}%
  \!ucoord{\number\y}{\number\yl}{\number\y}{\number\auxa}{\number\d}{\yb}{1}%
  \xmed=\xa%
  \advance \xmed by \xb%
  \divide \xmed by 2
  \ymed=\ya%
  \advance \ymed by \yb%
  \divide \ymed by 2
  \!distnomemor{\number\x}{\number\y}{\number\xl}{\number\yl}{#7}{\dnm}%
  \!ucoord{\number\y}{\number\yl}{\number\xmed}{\number\dnm}{\number\d}{\xc}{-#8}%
  \!ucoord{\number\x}{\number\xl}{\number\ymed}{\number\dnm}{\number\d}{\yc}{#8}%
\ifcase #9  
  \arrow <4pt> [.2,1.1] from {\xa} {\ya} to {\xb} {\yb}
\or  
  \setdashes
  \arrow <4pt> [.2,1.1] from {\xa} {\ya} to {\xb} {\yb}
  \setsolid
\or  
  \setlinear
  \plot {\xa} {\ya}  {\xb} {\yb} /
\or  
  \auxa=\guardaauxa
  \advance \auxa by 3%
 \!ucoord{\number\x}{\number\xl}{\number\x}{\number\auxa}{\number\d}{\xa}{1}%
 \!ucoord{\number\y}{\number\yl}{\number\y}{\number\auxa}{\number\d}{\ya}{1}%
 \!ucoord{\number\y}{\number\yl}{\number\xa}{3}{\number\d}{\xd}{-1}%
 \!ucoord{\number\x}{\number\xl}{\number\ya}{3}{\number\d}{\yd}{1}%
  \arrow <4pt> [.2,1.1] from {\xa} {\ya} to {\xb} {\yb}
  \circulararc -180 degrees from {\xa} {\ya} center at {\xd} {\yd}
\or  
  \auxa=3
 \!ucoord{\number\y}{\number\yl}{\number\xa}{\number\auxa}{\number\d}{\xmed}{-1}%
 \!ucoord{\number\x}{\number\xl}{\number\ya}{\number\auxa}{\number\d}{\ymed}{1}%
 \!ucoord{\number\y}{\number\yl}{\number\xa}{\number\auxa}{\number\d}{\xd}{1}%
 \!ucoord{\number\x}{\number\xl}{\number\ya}{\number\auxa}{\number\d}{\yd}{-1}%
  \arrow <4pt> [.2,1.1] from {\xa} {\ya} to {\xb} {\yb}
  \setlinear
  \plot {\xmed} {\ymed}  {\xd} {\yd} /
\fi
\auxa=\xl
\advance \auxa by -\x%
\ifnum \auxa=0 
  \put {#7} at {\xc} {\yc}
\else
  \auxb=\yl
  \advance \auxb by -\y%
  \ifnum \auxb=0 \put {#7} at {\xc} {\yc}
  \else 
    \ifnum \auxa > 0 
      \ifnum \auxb > 0
        \ifnum #8=1
          \put {#7} [rb] at {\xc} {\yc}
        \else 
          \put {#7} [lt] at {\xc} {\yc}
        \fi
      \else
        \ifnum #8=1
          \put {#7} [lb] at {\xc} {\yc}
        \else 
          \put {#7} [rt] at {\xc} {\yc}
        \fi
      \fi
    \else
      \ifnum \auxb > 0 
        \ifnum #8=1
          \put {#7} [rt] at {\xc} {\yc}
        \else 
          \put {#7} [lb] at {\xc} {\yc}
        \fi
      \else
        \ifnum #8=1
          \put {#7} [lt] at {\xc} {\yc}
        \else 
          \put {#7} [rb] at {\xc} {\yc}
        \fi
      \fi
    \fi
  \fi
\fi
}

\def\modifplot(#1{\!modifqcurve #1}
\def\!modifqcurve(#1,#2){\x=#1%
  \y=#2%
  \multiply \x by \expansao%
  \multiply \y by \expansao%
  \!start (\x,\y)
  \!modifQjoin}
\def\!modifQjoin(#1,#2)(#3,#4){\x=#1%
  \y=#2%
  \xl=#3%
  \yl=#4%
  \multiply \x by \expansao%
  \multiply \y by \expansao%
  \multiply \xl by \expansao%
  \multiply \yl by \expansao%
  \!qjoin (\x,\y) (\xl,\yl)             
  \!ifnextchar){\!fim}{\!modifQjoin}}
\def\!fim){\ignorespaces}

\def\setaxy(#1{\!pontosxy #1}
\def\!pontosxy(#1,#2){%
  \!maispontosxy}
\def\!maispontosxy(#1,#2)(#3,#4){%
  \!ifnextchar){\!fimxy#3,#4}{\!maispontosxy}}
\def\!fimxy#1,#2){\x=#1%
  \y=#2
  \multiply \x by \expansao
  \multiply \y by \expansao
  \xl=\x%
  \yl=\y%
  \aux=1%
  \multiply \aux by \auxa%
  \advance\xl by \aux%
  \aux=1%
  \multiply \aux by \auxb%
  \advance\yl by \aux%
  \arrow <4pt> [.2,1.1] from {\x} {\y} to {\xl} {\yl}}

\def\cmor#1 #2(#3,#4)#5{%
  \!ifnextchar[{\!cmora{#1}{#2}{#3}{#4}{#5}}{\!cmora{#1}{#2}{#3}{#4}{#5}[0] }}
\def\!cmora#1#2#3#4#5[#6]{%
  \ifcase #2
      \auxa=0
      \auxb=1
    \or
      \auxa=0
      \auxb=-1
    \or
      \auxa=1
      \auxb=0
    \or
      \auxa=-1
      \auxb=0
    \fi  
  \ifcase #6  
    \modifplot#1
    \setaxy#1
  \or  
    \setdashes
    \modifplot#1
    \setaxy#1
    \setsolid
  \or  
    \modifplot#1
  \fi  
  \x=#3%
  \y=#4%
  \multiply \x by \expansao%
  \multiply \y by \expansao%
  \put {#5} at {\x} {\y}}

\def\obj(#1,#2){%
  \!ifnextchar[{\!obja{#1}{#2}}{\!obja{#1}{#2}[Nulo]}}
\def\!obja#1#2[#3]#4{%
  \!ifnextchar[{\!objb{#1}{#2}{#3}{#4}}{\!objb{#1}{#2}{#3}{#4}[1]}}
\def\!objb#1#2#3#4[#5]{%
  \x=#1%
  \y=#2%
  \def\!pinta{\normalsize$\bullet$}
  \def\!nulo{Nulo}%
  \def\!arg{#3}%
  \!compara{\!arg}{\!nulo}%
  \ifcompara\def\!arg{#4}\fi%
  \multiply \x by \expansao%
  \multiply \y by \expansao%
  \setbox\caixa=\hbox{#4}%
  \!coloca{(\!arg)(#1,#2)(\number\ht\caixa,\number\wd\caixa,\number\dp\caixa)}{\pilha}%
  \auxa=\wd\caixa \divide \auxa by 131072 
  \advance \auxa by 5
  \auxb=\ht\caixa
  \advance \auxb by \number\dp\caixa
  \divide \auxb by 131072 
  \advance \auxb by 5
  \ifcase \tipografo    
    \put{#4} at {\x} {\y}
  \or                   
    \ifcase #5 
      \put{#4} at {\x} {\y}
    \or        
      \put{\!pinta} at {\x} {\y}
      \advance \y by \number\auxb  
      \put{#4} at {\x} {\y}
    \or        
      \put{\!pinta} at {\x} {\y}
      \advance \auxa by -2  
      \advance \auxb by -2  
      \advance \x by \number\auxa  
      \advance \y by \number\auxb  
      \put{#4} at {\x} {\y}   
    \or        
      \put{\!pinta} at {\x} {\y}
      \advance \x by \number\auxa  
      \put{#4} at {\x} {\y}   
    \or        
      \put{\!pinta} at {\x} {\y}
      \advance \auxa by -2  
      \advance \auxb by -2  
      \advance \x by \number\auxa  
      \advance \y by -\number\auxb  
      \put{#4} at {\x} {\y}   
    \or        
      \put{\!pinta} at {\x} {\y}
      \advance \y by -\number\auxb  
      \put{#4} at {\x} {\y}   
    \or        
      \put{\!pinta} at {\x} {\y}
      \advance \auxa by -2  
      \advance \auxb by -2  
      \advance \x by -\number\auxa  
      \advance \y by -\number\auxb  
      \put{#4} at {\x} {\y}   
    \or        
      \put{\!pinta} at {\x} {\y}
      \advance \x by -\number\auxa  
      \put{#4} at {\x} {\y}   
    \or        
      \put{\!pinta} at {\x} {\y}
      \advance \auxa by -2  
      \advance \auxb by -2  
      \advance \x by -\number\auxa  
      \advance \y by \number\auxb  
      \put{#4} at {\x} {\y}   
    \fi
  \or                   
    \ifcase #5 
      \put{#4} at {\x} {\y}
    \or        
      \put{\!pinta} at {\x} {\y}
      \advance \y by \number\auxb  
      \put{#4} at {\x} {\y}
    \or        
      \put{\!pinta} at {\x} {\y}
      \advance \auxa by -2  
      \advance \auxb by -2  
      \advance \x by \number\auxa  
      \advance \y by \number\auxb  
      \put{#4} at {\x} {\y}   
    \or        
      \put{\!pinta} at {\x} {\y}
      \advance \x by \number\auxa  
      \put{#4} at {\x} {\y}   
    \or        
      \put{\!pinta} at {\x} {\y}
      \advance \auxa by -2  
      \advance \auxb by -2
      \advance \x by \number\auxa  
      \advance \y by -\number\auxb 
      \put{#4} at {\x} {\y}   
    \or        
      \put{\!pinta} at {\x} {\y}
      \advance \y by -\number\auxb 
      \put{#4} at {\x} {\y}   
    \or        
      \put{\!pinta} at {\x} {\y}
      \advance \auxa by -2  
      \advance \auxb by -2
      \advance \x by -\number\auxa 
      \advance \y by -\number\auxb 
      \put{#4} at {\x} {\y}   
    \or        
      \put{\!pinta} at {\x} {\y}
      \advance \x by -\number\auxa 
      \put{#4} at {\x} {\y}   
    \or        
      \put{\!pinta} at {\x} {\y}
      \advance \auxa by -2  
      \advance \auxb by -2
      \advance \x by -\number\auxa 
      \advance \y by \number\auxb  
      \put{#4} at {\x} {\y}   
    \fi
   \else 
     \ifnum\auxa<\auxb 
       \aux=\auxb
     \else
       \aux=\auxa
     \fi
     \ifdim\wd\caixa<1em
       \dimen99 = 1em
       \aux=\dimen99 \divide \aux by 131072 
       \advance \aux by 5
     \fi
     \advance\aux by -2 
     \multiply\aux by 2 %
     \ifnum\aux<30
       \put{\circle{\aux}} [Bl] at {\x} {\y}
     \else
       \multiply\auxa by 2
       \multiply\auxb by 2
       \put{\oval(\auxa,\auxb)} [Bl] at {\x} {\y}
     \fi
     \put{#4} at {\x} {\y}
   \fi   
}

\catcode`!=12 

%
%

%
%
%
%

\def\Serif{cmr}
\def\SerifBold{cmbx}
\def\SerifItalics{cmti}
\def\SerifSlanted{cmsl}
\def\SerifBoldItalics{cmbxti}
\def\SansSerif{cmss}
\def\SansSerifBold{cmssbx}
\def\SansSerifItalics{cmssi}
\def\SansSerifSlanted{cmssi}
\def\Math{cmmi}
\def\Symbols{cmsy}
\def\MathBold{cmmib}
\def\MoreSymbols{cmex}
\def\Typewriter{cmtt}
\def\Gothic{eufm}
\def\Double{msbm}

= 			\Serif10 			at 5pt
= 		\SerifBold10 		at 5pt
= 	\SerifItalics10 	at 5pt
=		\SerifSlanted10 	at 5pt
=	\SerifBoldItalics10	at 5pt
= 		\SansSerif10 		at 5pt
=	\SansSerifBold10	at 5pt
=	\SansSerifItalics10	at 5pt
=	\SansSerifSlanted10	at 5pt
=				\Math10				at 5pt
=			\MathBold10			at 5pt
=			\Symbols10			at 5pt
=		\MoreSymbols10		at 5pt
=		\Typewriter10		at 5pt
=			\Gothic10			at 5pt
=			\Double10			at 5pt

= 			\Serif10 			at 7pt
= 		\SerifBold10 		at 7pt
= 	\SerifItalics10 	at 7pt
=	\SerifSlanted10 	at 7pt
=\SerifBoldItalics10	at 7pt
= 		\SansSerif10 		at 7pt
= 	\SansSerifBold10 	at 7pt
=\SansSerifItalics10	at 7pt
=\SansSerifSlanted10	at 7pt
=			\Math10				at 7pt
=		\MathBold10			at 7pt
=			\Symbols10			at 7pt
=		\MoreSymbols10		at 7pt
=		\Typewriter10		at 7pt
=			\Gothic10			at 7pt
=			\Double10			at 7pt

= 			\Serif10 			at 8pt
= 		\SerifBold10 		at 8pt
= 	\SerifItalics10 	at 8pt
=	\SerifSlanted10 	at 8pt
=\SerifBoldItalics10	at 8pt
= 		\SansSerif10 		at 8pt
= 	\SansSerifBold10 	at 8pt
=\SansSerifItalics10 at 8pt
=\SansSerifSlanted10 at 8pt
=			\Math10				at 8pt
=		\MathBold10			at 8pt
=			\Symbols10			at 8pt
=		\MoreSymbols10		at 8pt
=		\Typewriter10		at 8pt
=			\Gothic10			at 8pt
=			\Double10			at 8pt

= 			\Serif10 			at 10pt
= 		\SerifBold10 		at 10pt
= 		\SerifItalics10 	at 10pt
=		\SerifSlanted10 	at 10pt
=	\SerifBoldItalics10	at 10pt
= 		\SansSerif10 		at 10pt
= 	\SansSerifBold10 	at 10pt
= 	\SansSerifItalics10 at 10pt
= 	\SansSerifSlanted10 at 10pt
=				\Math10				at 10pt
=			\MathBold10			at 10pt
=			\Symbols10			at 10pt
=		\MoreSymbols10		at 10pt
=		\Typewriter10		at 10pt
=			\Gothic10			at 10pt
=			\Double10			at 10pt

= 				\Serif10 			at 12pt
= 			\SerifBold10 		at 12pt
= 		\SerifItalics10 	at 12pt
=		\SerifSlanted10 	at 12pt
=	\SerifBoldItalics10	at 12pt
= 			\SansSerif10 		at 12pt
= 		\SansSerifBold10 	at 12pt
= 	\SansSerifItalics10 at 12pt
= 	\SansSerifSlanted10 at 12pt
=				\Math10				at 12pt
=			\MathBold10			at 12pt
=			\Symbols10			at 12pt
=		\MoreSymbols10		at 12pt
=			\Typewriter10		at 12pt
=				\Gothic10			at 12pt
=				\Double10			at 12pt

= 			\Serif10 			at 14pt
= 		\SerifBold10 		at 14pt
= 	\SerifItalics10 	at 14pt
=		\SerifSlanted10 	at 14pt
=	\SerifBoldItalics10	at 14pt
= 		\SansSerif10 		at 14pt
= 	\SansSerifBold10 	at 14pt
= \SansSerifSlanted10 at 14pt
= \SansSerifItalics10 at 14pt
=				\Math10				at 14pt
=			\MathBold10			at 14pt
=			\Symbols10			at 14pt
=		\MoreSymbols10		at 14pt
=		\Typewriter10		at 14pt
=			\Gothic10			at 14pt
=			\Double10			at 14pt

\def\NormalStyle{\parindent=5pt\parskip=3pt\normalbaselineskip=14pt%
\def\nt{\tenSerif}%
\def\rm{\fam0\tenSerif}%
\textfont0=\tenSerif\scriptfont0=\sevenSerif\scriptscriptfont0=\fiveSerif
\textfont1=\tenMath\scriptfont1=\sevenMath\scriptscriptfont1=\fiveMath
\textfont2=\tenSymbols\scriptfont2=\sevenSymbols\scriptscriptfont2=\fiveSymbols
\textfont3=\tenMoreSymbols\scriptfont3=\sevenMoreSymbols\scriptscriptfont3=\fiveMoreSymbols
\textfont\itfam=\tenSerifItalics\def\it{\fam\itfam\tenSerifItalics}%
\textfont\slfam=\tenSerifSlanted\def\sl{\fam\slfam\tenSerifSlanted}%
\textfont\ttfam=\tenTypewriter\def\tt{\fam\ttfam\tenTypewriter}%
\textfont\bffam=\tenSerifBold%
\def\bf{\fam\bffam\tenSerifBold}\scriptfont\bffam=\sevenSerifBold\scriptscriptfont\bffam=\fiveSerifBold%
\def\cal{\tenSymbols}%
\def\greekbold{\tenMathBold}%
\def\gothic{\tenGothic}%
\def\Bbb{\tenDouble}%
\def\LieFont{\tenSerifItalics}%
\nt\normalbaselines\baselineskip=14pt%
}

\def\TitleStyle{\parindent=0pt\parskip=0pt\normalbaselineskip=15pt%
\def\nt{\fourteenSansSerifBold}%
\def\rm{\fam0\fourteenSansSerifBold}%
\textfont0=\fourteenSansSerifBold\scriptfont0=\tenSansSerifBold\scriptscriptfont0=\eightSansSerifBold
\textfont1=\fourteenMath\scriptfont1=\tenMath\scriptscriptfont1=\eightMath
\textfont2=\fourteenSymbols\scriptfont2=\tenSymbols\scriptscriptfont2=\eightSymbols
\textfont3=\fourteenMoreSymbols\scriptfont3=\tenMoreSymbols\scriptscriptfont3=\eightMoreSymbols
\textfont\itfam=\fourteenSansSerifItalics\def\it{\fam\itfam\fourteenSansSerifItalics}%
\textfont\slfam=\fourteenSansSerifSlanted\def\sl{\fam\slfam\fourteenSerifSansSlanted}%
\textfont\ttfam=\fourteenTypewriter\def\tt{\fam\ttfam\fourteenTypewriter}%
\textfont\bffam=\fourteenSansSerif%
\def\bf{\fam\bffam\fourteenSansSerif}\scriptfont\bffam=\tenSansSerif\scriptscriptfont\bffam=\eightSansSerif%
\def\cal{\fourteenSymbols}%
\def\greekbold{\fourteenMathBold}%
\def\gothic{\fourteenGothic}%
\def\Bbb{\fourteenDouble}%
\def\LieFont{\fourteenSerifItalics}%
\nt\normalbaselines\baselineskip=15pt%
}

\def\PartStyle{\parindent=0pt\parskip=0pt\normalbaselineskip=15pt%
\def\nt{\fourteenSansSerifBold}%
\def\rm{\fam0\fourteenSansSerifBold}%
\textfont0=\fourteenSansSerifBold\scriptfont0=\tenSansSerifBold\scriptscriptfont0=\eightSansSerifBold
\textfont1=\fourteenMath\scriptfont1=\tenMath\scriptscriptfont1=\eightMath
\textfont2=\fourteenSymbols\scriptfont2=\tenSymbols\scriptscriptfont2=\eightSymbols
\textfont3=\fourteenMoreSymbols\scriptfont3=\tenMoreSymbols\scriptscriptfont3=\eightMoreSymbols
\textfont\itfam=\fourteenSansSerifItalics\def\it{\fam\itfam\fourteenSansSerifItalics}%
\textfont\slfam=\fourteenSansSerifSlanted\def\sl{\fam\slfam\fourteenSerifSansSlanted}%
\textfont\ttfam=\fourteenTypewriter\def\tt{\fam\ttfam\fourteenTypewriter}%
\textfont\bffam=\fourteenSansSerif%
\def\bf{\fam\bffam\fourteenSansSerif}\scriptfont\bffam=\tenSansSerif\scriptscriptfont\bffam=\eightSansSerif%
\def\cal{\fourteenSymbols}%
\def\greekbold{\fourteenMathBold}%
\def\gothic{\fourteenGothic}%
\def\Bbb{\fourteenDouble}%
\def\LieFont{\fourteenSerifItalics}%
\nt\normalbaselines\baselineskip=15pt%
}

\def\ChapterStyle{\parindent=0pt\parskip=0pt\normalbaselineskip=15pt%
\def\nt{\fourteenSansSerifBold}%
\def\rm{\fam0\fourteenSansSerifBold}%
\textfont0=\fourteenSansSerifBold\scriptfont0=\tenSansSerifBold\scriptscriptfont0=\eightSansSerifBold
\textfont1=\fourteenMath\scriptfont1=\tenMath\scriptscriptfont1=\eightMath
\textfont2=\fourteenSymbols\scriptfont2=\tenSymbols\scriptscriptfont2=\eightSymbols
\textfont3=\fourteenMoreSymbols\scriptfont3=\tenMoreSymbols\scriptscriptfont3=\eightMoreSymbols
\textfont\itfam=\fourteenSansSerifItalics\def\it{\fam\itfam\fourteenSansSerifItalics}%
\textfont\slfam=\fourteenSansSerifSlanted\def\sl{\fam\slfam\fourteenSerifSansSlanted}%
\textfont\ttfam=\fourteenTypewriter\def\tt{\fam\ttfam\fourteenTypewriter}%
\textfont\bffam=\fourteenSansSerif%
\def\bf{\fam\bffam\fourteenSansSerif}\scriptfont\bffam=\tenSansSerif\scriptscriptfont\bffam=\eightSansSerif%
\def\cal{\fourteenSymbols}%
\def\greekbold{\fourteenMathBold}%
\def\gothic{\fourteenGothic}%
\def\Bbb{\fourteenDouble}%
\def\LieFont{\fourteenSerifItalics}%
\nt\normalbaselines\baselineskip=15pt%
}

\def\SectionStyle{\parindent=0pt\parskip=0pt\normalbaselineskip=13pt%
\def\nt{\twelveSansSerifBold}%
\def\rm{\fam0\twelveSansSerifBold}%
\textfont0=\twelveSansSerifBold\scriptfont0=\eightSansSerifBold\scriptscriptfont0=\eightSansSerifBold
\textfont1=\twelveMath\scriptfont1=\eightMath\scriptscriptfont1=\eightMath
\textfont2=\twelveSymbols\scriptfont2=\eightSymbols\scriptscriptfont2=\eightSymbols
\textfont3=\twelveMoreSymbols\scriptfont3=\eightMoreSymbols\scriptscriptfont3=\eightMoreSymbols
\textfont\itfam=\twelveSansSerifItalics\def\it{\fam\itfam\twelveSansSerifItalics}%
\textfont\slfam=\twelveSansSerifSlanted\def\sl{\fam\slfam\twelveSerifSansSlanted}%
\textfont\ttfam=\twelveTypewriter\def\tt{\fam\ttfam\twelveTypewriter}%
\textfont\bffam=\twelveSansSerif%
\def\bf{\fam\bffam\twelveSansSerif}\scriptfont\bffam=\eightSansSerif\scriptscriptfont\bffam=\eightSansSerif%
\def\cal{\twelveSymbols}%
\def\bg{\twelveMathBold}%
\def\gothic{\twelveGothic}%
\def\Bbb{\twelveDouble}%
\def\LieFont{\twelveSerifItalics}%
\nt\normalbaselines\baselineskip=13pt%
}

\def\SubSectionStyle{\parindent=0pt\parskip=0pt\normalbaselineskip=13pt%
\def\nt{\twelveSansSerifItalics}%
\def\rm{\fam0\twelveSansSerifItalics}%
\textfont0=\twelveSansSerifItalics\scriptfont0=\eightSansSerifItalics\scriptscriptfont0=\eightSansSerifItalics%
\textfont1=\twelveMath\scriptfont1=\eightMath\scriptscriptfont1=\eightMath%
\textfont2=\twelveSymbols\scriptfont2=\eightSymbols\scriptscriptfont2=\eightSymbols%
\textfont3=\twelveMoreSymbols\scriptfont3=\eightMoreSymbols\scriptscriptfont3=\eightMoreSymbols%
\textfont\itfam=\twelveSansSerif\def\it{\fam\itfam\twelveSansSerif}%
\textfont\slfam=\twelveSansSerifSlanted\def\sl{\fam\slfam\twelveSerifSansSlanted}%
\textfont\ttfam=\twelveTypewriter\def\tt{\fam\ttfam\twelveTypewriter}%
\textfont\bffam=\twelveSansSerifBold%
\def\bf{\fam\bffam\twelveSansSerifBold}\scriptfont\bffam=\eightSansSerifBold\scriptscriptfont\bffam=\eightSansSerifBold%
\def\cal{\twelveSymbols}%
\def\greekbold{\twelveMathBold}%
\def\gothic{\twelveGothic}%
\def\Bbb{\twelveDouble}%
\def\LieFont{\twelveSerifItalics}%
\nt\normalbaselines\baselineskip=13pt%
}

\def\AuthorStyle{\parindent=0pt\parskip=0pt\normalbaselineskip=14pt%
\def\nt{\tenSerif}%
\def\rm{\fam0\tenSerif}%
\textfont0=\tenSerif\scriptfont0=\sevenSerif\scriptscriptfont0=\fiveSerif
\textfont1=\tenMath\scriptfont1=\sevenMath\scriptscriptfont1=\fiveMath
\textfont2=\tenSymbols\scriptfont2=\sevenSymbols\scriptscriptfont2=\fiveSymbols
\textfont3=\tenMoreSymbols\scriptfont3=\sevenMoreSymbols\scriptscriptfont3=\fiveMoreSymbols
\textfont\itfam=\tenSerifItalics\def\it{\fam\itfam\tenSerifItalics}%
\textfont\slfam=\tenSerifSlanted\def\sl{\fam\slfam\tenSerifSlanted}%
\textfont\ttfam=\tenTypewriter\def\tt{\fam\ttfam\tenTypewriter}%
\textfont\bffam=\tenSerifBold%
\def\bf{\fam\bffam\tenSerifBold}\scriptfont\bffam=\sevenSerifBold\scriptscriptfont\bffam=\fiveSerifBold%
\def\cal{\tenSymbols}%
\def\greekbold{\tenMathBold}%
\def\gothic{\tenGothic}%
\def\Bbb{\tenDouble}%
\def\LieFont{\tenSerifItalics}%
\nt\normalbaselines\baselineskip=14pt%
}

\def\AddressStyle{\parindent=0pt\parskip=0pt\normalbaselineskip=14pt%
\def\nt{\eightSerif}%
\def\rm{\fam0\eightSerif}%
\textfont0=\eightSerif\scriptfont0=\sevenSerif\scriptscriptfont0=\fiveSerif
\textfont1=\eightMath\scriptfont1=\sevenMath\scriptscriptfont1=\fiveMath
\textfont2=\eightSymbols\scriptfont2=\sevenSymbols\scriptscriptfont2=\fiveSymbols
\textfont3=\eightMoreSymbols\scriptfont3=\sevenMoreSymbols\scriptscriptfont3=\fiveMoreSymbols
\textfont\itfam=\eightSerifItalics\def\it{\fam\itfam\eightSerifItalics}%
\textfont\slfam=\eightSerifSlanted\def\sl{\fam\slfam\eightSerifSlanted}%
\textfont\ttfam=\eightTypewriter\def\tt{\fam\ttfam\eightTypewriter}%
\textfont\bffam=\eightSerifBold%
\def\bf{\fam\bffam\eightSerifBold}\scriptfont\bffam=\sevenSerifBold\scriptscriptfont\bffam=\fiveSerifBold%
\def\cal{\eightSymbols}%
\def\greekbold{\eightMathBold}%
\def\gothic{\eightGothic}%
\def\Bbb{\eightDouble}%
\def\LieFont{\eightSerifItalics}%
\nt\normalbaselines\baselineskip=14pt%
}

\def\AbstractStyle{\parindent=0pt\parskip=0pt\normalbaselineskip=12pt%
\def\nt{\eightSerif}%
\def\rm{\fam0\eightSerif}%
\textfont0=\eightSerif\scriptfont0=\sevenSerif\scriptscriptfont0=\fiveSerif
\textfont1=\eightMath\scriptfont1=\sevenMath\scriptscriptfont1=\fiveMath
\textfont2=\eightSymbols\scriptfont2=\sevenSymbols\scriptscriptfont2=\fiveSymbols
\textfont3=\eightMoreSymbols\scriptfont3=\sevenMoreSymbols\scriptscriptfont3=\fiveMoreSymbols
\textfont\itfam=\eightSerifItalics\def\it{\fam\itfam\eightSerifItalics}%
\textfont\slfam=\eightSerifSlanted\def\sl{\fam\slfam\eightSerifSlanted}%
\textfont\ttfam=\eightTypewriter\def\tt{\fam\ttfam\eightTypewriter}%
\textfont\bffam=\eightSerifBold%
\def\bf{\fam\bffam\eightSerifBold}\scriptfont\bffam=\sevenSerifBold\scriptscriptfont\bffam=\fiveSerifBold%
\def\cal{\eightSymbols}%
\def\greekbold{\eightMathBold}%
\def\gothic{\eightGothic}%
\def\Bbb{\eightDouble}%
\def\LieFont{\eightSerifItalics}%
\nt\normalbaselines\baselineskip=12pt%
}

\def\RefsStyle{\parindent=0pt\parskip=0pt%
\def\nt{\eightSerif}%
\def\rm{\fam0\eightSerif}%
\textfont0=\eightSerif\scriptfont0=\sevenSerif\scriptscriptfont0=\fiveSerif
\textfont1=\eightMath\scriptfont1=\sevenMath\scriptscriptfont1=\fiveMath
\textfont2=\eightSymbols\scriptfont2=\sevenSymbols\scriptscriptfont2=\fiveSymbols
\textfont3=\eightMoreSymbols\scriptfont3=\sevenMoreSymbols\scriptscriptfont3=\fiveMoreSymbols
\textfont\itfam=\eightSerifItalics\def\it{\fam\itfam\eightSerifItalics}%
\textfont\slfam=\eightSerifSlanted\def\sl{\fam\slfam\eightSerifSlanted}%
\textfont\ttfam=\eightTypewriter\def\tt{\fam\ttfam\eightTypewriter}%
\textfont\bffam=\eightSerifBold%
\def\bf{\fam\bffam\eightSerifBold}\scriptfont\bffam=\sevenSerifBold\scriptscriptfont\bffam=\fiveSerifBold%
\def\cal{\eightSymbols}%
\def\greekbold{\eightMathBold}%
\def\gothic{\eightGothic}%
\def\Bbb{\eightDouble}%
\def\LieFont{\eightSerifItalics}%
\nt\normalbaselines\baselineskip=10pt%
}



%
%


\def\ModeYes{yes}
\def\ModeNo{no}

\def\ModeUndef{undefined}


\def\nx{\noexpand}
\def\ni{\noindent}
\def\newpage{\vfill\eject}

\def\ss{\vskip 5pt}
\def\ms{\vskip 10pt}
\def\bs{\vskip 20pt}

 \def\,{\mskip\thinmuskip}
 \def\!{\mskip-\thinmuskip}
 \def\>{\mskip\medmuskip}
 \def\;{\mskip\thickmuskip}

%
% ... 
%

\def\refsModePost{post}
\def\refsModeAuto{auto}

\def\dbRefsSatusModeOk{ok}
\def\dbRefsSatusModeError{error}
\def\dbRefsSatusModeWarning{warning}


\newcount\BNUM
\BNUM=0

\def\refs{}

\def\SetModePost{\xdef\refsMode{\refsModePost}}			
\SetModePost

\def\dbRefsStatusOk{%
	\xdef\dbRefsStatus{\dbRefsSatusModeOk}%
	\xdef\dbRefsError{\ModeNo}%
	\xdef\dbRefsWarning{\ModeNo}%
	\xdef\dbRefsInfo{\ModeNo}%
}

\def\dbRefs{%
}

\def\dbRefsGet#1{%
	\xdef\found{N}\xdef\ikey{#1}\dbRefsStatusOk%
	\xdef\key{\ModeUndef}\xdef\tag{\ModeUndef}\xdef\tail{\ModeUndef}%
	\dbRefs%
}

\def\NextRefsTag{%
	\global\advance\BNUM by 1%
}
\def\ShowTag#1{{\bf [#1]}}

\def\dbRefsInsert#1#2{%
\dbRefsGet{#1}%
\if\found Y %
   \xdef\dbRefsStatus{\dbRefsSatusModeWarning}%
   \xdef\dbRefsWarning{record is already there}%
   \xdef\dbRefsInfo{record not inserted}%
\else%
   \toks2=\expandafter{\dbRefs}%
   \ifx\refsMode\refsModeAuto \NextRefsTag
    \xdef\dbRefs{%
   	\the\toks2 \nx\xdef\nx\dbx{#1}%
	\nx\ifx\nx\ikey %
		\nx\dbx\nx\xdef\nx\found{Y}%
		\nx\xdef\nx\key{#1}%
		\nx\xdef\nx\tag{\the\BNUM}%
		\nx\xdef\nx\tail{#2}%
	\nx\fi}%
	\global\xdef\refs{\refs \ss\ni[\the\BNUM]\ #2\par}
   \fi%
   \ifx\refsMode\refsModePost 
    \xdef\dbRefs{%
   	\the\toks2 \nx\xdef\nx\dbx{#1}%
	\nx\ifx\nx\ikey %
		\nx\dbx\nx\xdef\nx\found{Y}%
		\nx\xdef\nx\key{#1}%
		\nx\xdef\nx\tag{\ModeUndef}%
		\nx\xdef\nx\tail{#2}%
	\nx\fi}%
   \fi%
\fi%
}

\def\dbRefsEdit#1#2#3{\dbRefsGet{#1}%
\if\found N 
   \xdef\dbRefsStatus{\dbRefsSatusModeError}%
   \xdef\dbRefsError{record is not there}%
   \xdef\dbRefsInfo{record not edited}%
\else%
   \toks2=\expandafter{\dbRefs}%
   \xdef\dbRefs{\the\toks2%
   \nx\xdef\nx\dbx{#1}%
   \nx\ifx\nx\ikey\nx\dbx %
	\nx\xdef\nx\found{Y}%
	\nx\xdef\nx\key{#1}%
	\nx\xdef\nx\tag{#2}%
	\nx\xdef\nx\tail{#3}%
   \nx\fi}%
\fi%
}

\def\bib#1#2{\RefsStyle\dbRefsInsert{#1}{#2}%
	\ifx\dbRefsStatus\dbRefsSatusModeWarning %
		\message{^^J}%
		\message{WARNING: Reference [#1] is doubled.^^J}%
	\fi%
}

\def\ref#1{\dbRefsGet{#1}%
\ifx\found N %
  \message{^^J}%
  \message{ERROR: Reference [#1] unknown.^^J}%
  \ShowTag{??}%
\else%
	\ifx\tag\ModeUndef \NextRefsTag%
		\dbRefsEdit{#1}{\the\BNUM}{\tail}%
		\dbRefsGet{#1}%
		\global\xdef\refs{\refs \ss\ni [\tag]\ \tail\par}
	\fi
	\ShowTag{\tag}%
\fi%
}

\def\ShowBiblio{\bs\Ensure{\SectionEnsure}%
{\SectionStyle\ni References}%
{\RefsStyle\refs}%
}

\newcount\CHANGES
\CHANGES=0
\def\AuxFile{7}
\def\PreventDoubleOn{\xdef\PreventDoubleLabel{\ModeYes}}

\PreventDoubleOn

\def\StoreLabel#1#2{\xdef\itag{#2}
 \ifx\PreModeStatus\ModeNo %
   \message{^^J}%
   \errmessage{You can't use Check without starting with OpenPreMode (and finishing with ClosePreMode)^^J}%
 \else%
   \immediate\write\AuxFile{\nx\dbLabelPreInsert{#1}{\itag}}%
   \dbLabelGet{#1}%
   \ifx\itag\tag %
   \else%
	\global\advance\CHANGES by 1%
 	\xdef\itag{(?.??)}%
    \fi%
   \fi%
}

\def\PreModeStatus{\ModeNo}

\def\ClosePreMode{\immediate\closeout\AuxFile%
  \ifnum\CHANGES=0%
	\message{^^J}%
	\message{**********************************^^J}%
	\message{**  NO CHANGES TO THE AuxFile  **^^J}%
	\message{**********************************^^J}%
 \else%
	\message{^^J}%
	\message{**************************************************^^J}%
	\message{**  PLAEASE TYPESET IT AGAIN (\the\CHANGES)  **^^J}%
    \errmessage{**************************************************^^ J}%
  \fi%
  \edef\PreModeStatus{\ModeNo}%
}

\def\dbLabelSatusModeOk{ok}

\def\dbLabelSatusModeWarning{warning}

\def\dbLabelStatusOk{%
	\xdef\dbLabelStatus{\dbLabelSatusModeOk}%
	\xdef\dbLabelError{\ModeNo}%
	\xdef\dbLabelWarning{\ModeNo}%
	\xdef\dbLabelInfo{\ModeNo}%
}

\def\dbLabel{%
}

\def\dbLabelGet#1{%
	\xdef\found{N}\xdef\ikey{#1}\dbLabelStatusOk%
	\xdef\key{\ModeUndef}\xdef\tag{\ModeUndef}\xdef\pre{\ModeUndef}%
	\dbLabel%
}

\def\ShowLabel#1{%
 \dbLabelGet{#1}%
 \ifx\tag \ModeUndef %
 	\global\advance\CHANGES by 1%
 	(?.??)%
 \else%
 	\tag%
 \fi%
}

\def\dbLabelPreInsert#1#2{\dbLabelGet{#1}%
\if\found Y %
  \xdef\dbLabelStatus{\dbLabelSatusModeWarning}%
   \xdef\dbLabelWarning{Label is already there}%
   \xdef\dbLabelInfo{Label not inserted}%
   \message{^^J}%
   \errmessage{Double pre definition of label [#1]^^J}%
\else%
   \toks2=\expandafter{\dbLabel}%
    \xdef\dbLabel{%
   	\the\toks2 \nx\xdef\nx\dbx{#1}%
	\nx\ifx\nx\ikey %
		\nx\dbx\nx\xdef\nx\found{Y}%
		\nx\xdef\nx\key{#1}%
		\nx\xdef\nx\tag{#2}%
		\nx\xdef\nx\pre{\ModeYes}%
	\nx\fi}%
\fi%
}

\def\dbLabelInsert#1#2{\dbLabelGet{#1}%
\xdef\itag{#2}%
\dbLabelGet{#1}%
\if\found Y %
	\ifx\tag\itag %
	\else%
	   \ifx\PreventDoubleLabel\ModeYes %
		\message{^^J}%
		\errmessage{Double definition of label [#1]^^J}%
	   \else%
		\message{^^J}%
		\message{Double definition of label [#1]^^J}%
	   \fi%
	\fi%
   \xdef\dbLabelStatus{\dbLabelSatusModeWarning}%
   \xdef\dbLabelWarning{Label is already there}%
   \xdef\dbLabelInfo{Label not inserted}%
\else%
   \toks2=\expandafter{\dbLabel}%
    \xdef\dbLabel{%
   	\the\toks2 \nx\xdef\nx\dbx{#1}%
	\nx\ifx\nx\ikey %
		\nx\dbx\nx\xdef\nx\found{Y}%
		\nx\xdef\nx\key{#1}%
		\nx\xdef\nx\tag{#2}%
		\nx\xdef\nx\pre{\ModeNo}%
	\nx\fi}%
\fi%
}


\newcount\PART
\newcount\CHAPTER
\newcount\SECTION
\newcount\SUBSECTION
\newcount\FNUMBER

\PART=0
\CHAPTER=0
\SECTION=0
\SUBSECTION=0	
\FNUMBER=0

\def\LastPart{\ModeUndef}
\def\LastChapter{\ModeUndef}
\def\LastSection{\ModeUndef}
\def\LastSubSection{\ModeUndef}
\def\LastClaim{\ModeUndef}
\def\Last{\ModeUndef}

\newdimen\TOBOTTOM
\newdimen\LIMIT

\def\Ensure#1{\ \par\ \immediate\LIMIT=#1\immediate\TOBOTTOM=\the\pagegoal\advance\TOBOTTOM by -\pagetotal%
\ifdim\TOBOTTOM<\LIMIT\newpage \else%
\vskip-\parskip\vskip-\parskip\vskip-\baselineskip\fi}

\def\PartLabel{\the\PART}
\def\NewPart#1{\global\advance\PART by 1%
         \bs\ni{\PartStyle  Part \PartLabel:}
         \bs\ni{\PartStyle #1}\newpage%
         \CHAPTER=0\SECTION=0\SUBSECTION=0\FNUMBER=0%
         \gdef\Left{#1}%
         \global\edef\Last{\PartLabel}%
         \global\edef\LastPart{\PartLabel}%
         \global\edef\LastChapter{\ModeUndef}%
         \global\edef\LastSection{\ModeUndef}%
         \global\edef\LastSubSection{\ModeUndef}%
         \global\edef\LastClaim{\ModeUndef}}
\def\ChapterLabel{\the\CHAPTER}
\def\NewChapter#1{\global\advance\CHAPTER by 1%
         \bs\ni{\ChapterStyle  Chapter \ChapterLabel: #1}\ms%
         \SECTION=0\SUBSECTION=0\FNUMBER=0%
         \gdef\Left{#1}%
         \global\edef\Last{\ChapterLabel}%
         \global\edef\LastChapter{\ChapterLabel}%
         \global\edef\LastSection{\ModeUndef}%
         \global\edef\LastSubSection{\ModeUndef}%
         \global\edef\LastClaim{\ModeUndef}}
\def\SectionEnsure{3cm}
\def\NewSection#1{\Ensure{\SectionEnsure}\gdef\SectionLabel{\the\SECTION}\global\advance\SECTION by 1%
         \bs\ni{\SectionStyle  \SectionLabel.\ #1}\ss%
         \SUBSECTION=0\FNUMBER=0%
         \gdef\Left{#1}%
         \global\edef\Last{\SectionLabel}%
         \global\edef\LastSection{\SectionLabel}%
         \global\edef\LastSubSection{\ModeUndef}%
         \global\edef\LastClaim{\ModeUndef}}
\def\NewAppendix#1#2{\Ensure{\SectionEnsure}\gdef\SectionLabel{#1}\global\advance\SECTION by 1%
         \bs\ni{\SectionStyle  Appendix \SectionLabel.\ #2}\ss%
         \SUBSECTION=0\FNUMBER=0%
         \gdef\Left{#2}%
         \global\edef\Last{\SectionLabel}%
         \global\edef\LastSection{\SectionLabel}%
         \global\edef\LastSubSection{\ModeUndef}%
         \global\edef\LastClaim{\ModeUndef}}
\def\Acknowledgements{\Ensure{\SectionEnsure}\gdef\SectionLabel{}%
         \bs\ni{\SectionStyle  Acknowledgments}\ss%
         \SECTION=0\SUBSECTION=0\FNUMBER=0%
         \gdef\Left{}%
         \global\edef\Last{\ModeUndef}%
         \global\edef\LastSection{\ModeUndef}%
         \global\edef\LastSubSection{\ModeUndef}%
         \global\edef\LastClaim{\ModeUndef}}
\def\SubSectionEnsure{2cm}
\def\SubSectionLabel{\ifnum\SECTION>0 \the\SECTION.\fi\the\SUBSECTION}
\def\NewSubSection#1{\Ensure{\SubSectionEnsure}\global\advance\SUBSECTION by 1%
         \ms\ni{\SubSectionStyle #1}\ss%
         \global\edef\Last{\SubSectionLabel}%
         \global\edef\LastSubSection{\SubSectionLabel}}
\def\SetNumberingModeN{\def\ClaimLabel{(\the\FNUMBER)}}
\def\SetNumberingModeSN{\def\ClaimLabel{(\ifnum\SECTION>0 \SectionLabel.\fi%
      \the\FNUMBER)}}
\def\SetNumberingModeCSN{\def\ClaimLabel{(\ifnum\CHAPTER>0 \the\CHAPTER.\fi%
      \ifnum\SECTION>0 \SectionLabel.\fi%
      \the\FNUMBER)}}

\def\NewClaim{\global\advance\FNUMBER by 1%
    \ClaimLabel%
    \global\edef\LastClaim{\ClaimLabel}%
    \global\edef\Last{\ClaimLabel}}

\def\HideLabels{\xdef\ShowLabelsMode{\ModeNo}}
\HideLabels

\def\fn{\eqno{\NewClaim}} 
\def\fl#1{%
\ifx\ShowLabelsMode\ModeYes%
 \eqno{{\buildrel{\hbox{\AbstractStyle[#1]}}\over{\hfill\NewClaim}}}%
\else%
 \eqno{\NewClaim}%
\fi%
\dbLabelInsert{#1}{\ClaimLabel}}
\def\fprel#1{\global\advance\FNUMBER by 1\StoreLabel{#1}{\ClaimLabel}%
\ifx\ShowLabelsMode\ModeYes%
\eqno{{\buildrel{\hbox{\AbstractStyle[#1]}}\over{\hfill.\itag}}}%
\else%
 \eqno{\itag}%
\fi%
}

\def\cl#1{\global\advance\FNUMBER by 1\dbLabelInsert{#1}{\ClaimLabel}%
\ifx\ShowLabelsMode\ModeYes%
${\buildrel{\hbox{\AbstractStyle[#1]}}\over{\hfill\ClaimLabel}}$%
\else%
  $\ClaimLabel$%
\fi%
}
\def\cprel#1{\global\advance\FNUMBER by 1\StoreLabel{#1}{\ClaimLabel}%
\ifx\ShowLabelsMode\ModeYes%
${\buildrel{\hbox{\AbstractStyle[#1]}}\over{\hfill.\itag}}$%
\else%
  $\itag$%
\fi%
}


\parindent=7pt
\leftskip=2cm
\newcount\SideIndent
\newcount\SideIndentTemp
\SideIndent=0
\newdimen\SectionIndent
\SectionIndent=-8pt

\def\sidebar{\vrule height15pt width.2pt }
\def\endcorner{\hbox{\hbox{\vrule height6pt width.2pt}\vbox to6pt{\vfill\hbox
to4pt{\leaders\hrule height0.2pt\hfill}}}}
\def\begincorner{\hbox{\hbox{\vrule height6pt width.2pt}\vbox to6pt{\hbox
to4pt{\leaders\hrule height0.2pt\hfill}}}}
\def\endbegincorner{\hbox{\vbox to15pt{\endcorner\vskip-6pt\begincorner\vfill}}}
\def\SideShow{\SideIndentTemp=\SideIndent \ifnum \SideIndentTemp>0 
\loop\sidebar\hskip 2pt \advance\SideIndentTemp by-1\ifnum \SideIndentTemp>1 \repeat\fi}

\def\BeginSection{{\vbadness 100000 \par\ni\hskip\SectionIndent%
\SideShow\vbox to 15pt{\vfill\begincorner}}\global\advance\SideIndent by1\vskip-10pt}

\def\EndSection{{\vbadness 100000 \par\ni\global\advance\SideIndent by-1%
\hskip\SectionIndent\SideShow\vbox to15pt{\endcorner\vfill}\vskip-10pt}}

\def\EndBeginSection{{\vbadness 100000\par\ni%
\global\advance\SideIndent by-1\hskip\SectionIndent\SideShow
\vbox to15pt{\vfill\endbegincorner}}%
\global\advance\SideIndent by1\vskip-10pt}

\def\ShowBeginCorners#1{%
\SideIndentTemp =#1 \advance\SideIndentTemp by-1%
\ifnum \SideIndentTemp>0 %
\vskip-15truept\hbox{\kern 2truept\vbox{\hbox{\begincorner}%
\ShowBeginCorners{\SideIndentTemp}\vskip-3truept}}%
\fi%
}

\def\ShowEndCorners#1{%
\SideIndentTemp =#1 \advance\SideIndentTemp by-1%
\ifnum \SideIndentTemp>0 %
\vskip-15truept\hbox{\kern 2truept\vbox{\hbox{\endcorner}%
\ShowEndCorners{\SideIndentTemp}\vskip 2truept}}%
\fi%
}

\def\BeginSections#1{{\vbadness 100000 \par\ni\hskip\SectionIndent%
\SideShow\vbox to 15pt{\vfill\ShowBeginCorners{#1}}}\global\advance\SideIndent by#1\vskip-10pt}

\def\EndSections#1{{\vbadness 100000 \par\ni\global\advance\SideIndent by-#1%
\hskip\SectionIndent\SideShow\vbox to15pt{\vskip15pt\ShowEndCorners{#1}\vfill}\vskip-10pt}}

\def\EndBeginSections#1#2{{\vbadness 100000\par\ni%
\global\advance\SideIndent by-#1%
\hbox{\hskip\SectionIndent\SideShow\kern-2pt%
\vbox to15pt{\vskip15pt\ShowEndCorners{#1}\vskip4pt\ShowBeginCorners{#2}}}}%
\global\advance\SideIndent by#2\vskip-10pt}




%
%


\def\ga{\gamma}

\def\ep{\epsilon}
\def\io{\iota}
\def\te{\theta}
\def\la{\lambda}

\def\om{\omega}
\def\si{\sigma}
\def\vp{\varphi}

\def\ka{\kappa}

\def\Ga{\Gamma}

\def\Si{\Sigma}


 \def\su{{\hbox{\gothic su}}}


 \def\one{{\hbox{\Bbb I}}}
 \def\A{{\hbox{\Bbb A}}}
 \def\R{{\hbox{\Bbb R}}}
 \def\C{{\hbox{\Bbb C}}}

 \def\R{{\hbox{\Bbb R}}}

 \def\K{{\hbox{\Bbb K}}}


\def\Aut{{\hbox{Aut}}}

\def\Spin{{\hbox{Spin}}}
\def\SO{{\hbox{SO}}}
\def\SU{{\hbox{SU}}}
\def\SL{{\hbox{SL}}}

\def\di{{\hbox{d}}}
\def\id{{\hbox{\rm id}}}

\def\ip{\hbox to4pt{\leaders\hrule height0.3pt\hfill}\vbox to8pt{\leaders\vrule width0.3pt\vfill}\kern 2pt}
 
\def\del{\partial}
\def\na{\nabla}

\def\arr{\rightarrow}
\def\harr{\hookrightarrow}

%
%

\def\cases#1{\left\{\eqalign{#1}\right.}
\NormalStyle
\SetNumberingModeSN
\PreventDoubleOn

\long\def\title#1{\centerline{\TitleStyle\ni#1}}
\long\def\author#1{\ms\centerline{\AuthorStyle by {\it #1}}}

\long\def\address#1{\ss\centerline{\AddressStyle #1}\par}
\long\def\moreaddress#1{\centerline{\AddressStyle #1}\par}
\def\abstract{\ms\leftskip 3cm\rightskip .5cm\AbstractStyle{\bf \ni Abstract:}\ }
\def\endabstract{\par\leftskip 2cm\rightskip 0cm\NormalStyle\ss}

\SetNumberingModeSN

\def\frac[#1/#2]{\hbox{$#1\over#2$}}

\def\({\left(}
\def\){\right)}
\def\[{\left[}
\def\]{\right]}
\def\^#1{{}^{#1}_{\>\cdot}}
\def\_#1{{}_{#1}^{\>\cdot}}
\def\Label=#1{{\buildrel {\hbox{\fiveSerif \ShowLabel{#1}}}\over =}}
\def\<{\kern -1pt}

\bib{OurAshtekar}{L.\ Fatibene, M.\ Francaviglia, {\it Spin Structures on Manifolds and Ashtekar Variables},Int. J. Geom. Methods Mod. Phys. {\bf 2}(2) (2005), 147-157}

\bib{Rovelli1}{L. Fatibene, M.Francaviglia, C.Rovelli, {\it On a Covariant Formulation of the Barberi-Immirzi Connection}
CQG 24 (2007) 3055-3066; gr-qc/0702134v1}

\bib{Rovelli2}{L. Fatibene, M.Francaviglia, C.Rovelli, {\it Spacetime Lagrangian Formulation of Barbero-Immirzi Gravity} 
CQG 24 (2007) 4207-4217; gr-qc/0706.1899}

\bib{Romani}{F. Cianfrani, O.M. Lecian, G. Montani, {\it Fundamentals and recent developments in non-perturbative canonical Quantum Gravity} ;  arXiv:0805.2503}

\bib{Book}{L.\ Fatibene, M.\ Francaviglia, {\it Natural and gauge natural formalism for classical field theories. A geometric perspective including spinors and gauge theories}, Kluwer Academic Publishers, Dordrecht, 2003}

\bib{Samuel}{J.\ Samuel, {\it Is Barbero's Hamiltonian Formulation a Gauge Theory of Lorentzian Gravity?}, Class.\ Quantum Grav.\ {\it 17}, 2000, 141-148}

\bib{Holst}{S.\ Holst, {\it Barbero's Hamiltonian Derived from a Generalized Hilbert-Palatini Action},
Phys.\ Rev.\ {\bf D53}, 5966, 1996}

\bib{ourHolst}{L.Fatibene, M.Ferraris, M.Francaviglia, G.Pacchiella, 
{\it Entropy of SelfÐGravitating Systems from HolstÕs Lagrangian}, 
Int. Journal of Geometrical Methods in Modern Physics,  {\bf 6}(2), 2009; gr-qc/0808.3845v2}

\bib{RovelliBook}{C.\ Rovelli, {\it Quantum Gravity}, Cambridge University Press, Cambridge, 2004}

\bib{RovelliBI}{A.\ Perez, C.\ Rovelli, {\it Physical effects of the Immirzi parameter},
Phys.\ Rev.\ {\bf D73}, 044013, 2006}

\bib{gr-qc/0404018}{A.\ Ashtekar, J.\ Lewandowski,
{\it Background Independent Quantum Gravity: a Status Report}, gr-qc/0404018}

\bib{Thiemann}{T.\ Thiemann, {\it Introduction to Modern Canonical Quantum general Relativity};
gr-qc/0110034}

\bib{Barbero}{F.\ Barbero, {\it Real Ashtekar variables for Lorentzian signature space-time},
Phys.\ Rev.\ {\it D51}, 5507, 1996}

\bib{Immirzi}{G.\ Immirzi, {\it Quantum Gravity and Regge Calculus},
Nucl.\ Phys.\ Proc.\ Suppl.\ {\bf 57}, 65-72}

\bib{KobaNu}{S.\ Kobayashi, K.\ Nomizu,
  {\it Foundations of differential geometry},
  John Wiley \& Sons, Inc., New York, 1963 USA}

\bib{Antonsen}{F.\ Antonsen, M.S.N.\ Flagga, {\it Spacetime Topology (I) - Chirality and the Third Stiefel-Whitney Class}, Int.\ J.\ Th.\ Phys.\ {\bf 41}(2), 2002}

\bib{Milnor}{J.W.\ Milnor, J.\ Stasheff, {\it Characteristic classes}, Princeton University Press, 1974}

\bib{Bombelli}{L.Bombelli, 
{Lagrangian Formulation}, in: {\it New Perspectives in Canonical Gravity},
Eds. A.Ashtekar, Bibliopolis (Napoli, 1988)}

\bib{Thie2006}{T. Thiemann,
{\it LoopQuantumGravity: An InsideView}, hep-th/0608210}

\NormalStyle
%

\title{Global Barbero-Immirzi Connections\footnote{$^*$}{{\AbstractStyle This contribution has been presented to the XVIII SIGRAV Conference, Cosenza (Italy), September 22-25, 2009.\goodbreak \ \goodbreak
This paper is published despite the effects of the Italian law 133/08 (see {\tt http://groups.google.it/group/scienceaction}). 
This law drastically reduces public funds to public Italian universities, which is particularly dangerous for free scientific research, 
and it will prevent young researchers from getting a position, either temporary or tenured, in Italy.
The authors are protesting against this law to obtain its cancellation.\goodbreak
}}}

\author{L.Fatibene$^{1,2}$, M.Francaviglia$^{1,2, 3}$}

\address{${ }^{1}$Dipartimento di Matematica, University of Torino}
\moreaddress{${ }^{2}$ INFN - Iniziativa Specifica Na12}
\moreaddress{${ }^{3}$ LCS - University of Cosenza}

\abstract
The Barbero-Immirzi (BI) connection, as usually introduced out of a spin connection, is a global object though it does not transform properly as a genuine connection with respect to generic spin transformations, unless quite specific and suitable gauges are imposed.
We shall here investigate whether and under which global conditions a (properly transforming and {\it hence} global) $\SU(2)$-connection can be canonically defined in a gauge covariant way in such a way that $\SU(2)$-connection locally agrees with the usual BI connection and can be defined on pretty general bundles (in particular triviality is not assumed).
As a by-product we shall also introduce a global covariant $\SU(2)$-connection over the whole spacetime (while for technical reasons  the BI connection in the standard formulation is just introduced on a space slice) which restricts to the usual BI connection on a space slice.
\endabstract

\NewSection{Introduction and Notation}

The Barbero-Immirzi (BI)  connection is introduced in LQG to describe the gravitational field by means of variables which are real also in Lorentz signature;
see   \ref{Barbero}, \ref{Immirzi}, \ref{RovelliBook}, \ref{gr-qc/0404018}, \ref{Thiemann} and references quoted therein.
Let us consider a $m=4$ dimensional spacetime $M$ and fix a signature $\eta$ which will be hereafter specialized to either the Euclidean case $\eta=(4,0)$ or to the Lorentzian case $\eta=(3,1)$.

If $\om^{ab}_\mu$ is the (4d) $\Spin(\eta)$-connection that is used in tetrad-affine formalism one first restricts it to some space leaf $i_t:S\harr M$
obtaining a (3d) $\Spin(\eta)$-connection $\om^{ab}_A$ on $S$. Latin indices from the firt part of the alphabet $a, b, \dots$ run from $0$ to $3$.

Instead of using this to define the selfdual connection, that would be complex in Lorentzian signature, one defines the BI field as
$$
A^k_A= \ga \om^{0k}_A+ \frac[1/2] \ep^k{}_{ij}\om^{ij}_A 
\fl{SpatialBIConnection}$$
where $\ga\in \R-\{0\}$ is called the Immirzi parameter. Latin indices $i,j,k,\dots$ run from $1$ to $3$.
We shall discuss below whether and in which sense the BI field can be regarded as a global connection. 

In Euclidean signature, for the special value $\ga=1$ (or $\ga=-1$) the BI connection so defined coincides with the (anti)-selfdual connection of $\om^{ab}_A$
and it is also the restriction to $S$ of the (anti)-selfdual connection defined on $M$ by $\om^{ab}_\mu$. 
For generic values of $\ga$ the definition of BI field \ShowLabel{SpatialBIConnection} corresponds to a canonical transformation.

Samuel (see \ref{Samuel}) provided an argument to claim that {\it the Barbero's connection cannot be interpreted as a spacetime connection}.
Of course it is difficult to precisely and rigorously determine what was exactly meant there by {\it spacetime interpretation}; while \ref{gr-qc/0404018} is more explicit in reporting Samuel's paper, claiming that it is impossible to obtain the BI connection as the restriction of a suitable global spacetime connection.
Also considering the details of their arguments we are forced to understand the claim is precisely in the sense that the BI connection is not  a restriction of a suitable globally defined spacetime $\Spin(3,1)$ connection.

We agree with Thiemann  (see \ref{Thie2006}) who refers to the problem as an {\it aesthetical} one, meaning that it would not spoil the mathematical consistency of the theory since the definition  \ShowLabel{SpatialBIConnection} is in any event a well-defined canonical transformation in the Hamiltonian theory defined on $S$. Nevertheless, we believe that a precise understanding of the geometric origin of fields is in any case needed, since it provides better insights on the structure of the theory.

We shall hereafter prove (see \ref{Rovelli1}, \ref{Rovelli2}) that the BI field defined above is in fact a connection and it is in fact the restriction of a spacetime connection, though not the restriction of a $\Spin(\eta)$ connection but rather the restriction of a $\SU(2)$-connection (let us remark that $\SU(2)\simeq \Spin(3)$).
The interpretation of such a (4d) $\SU(2)$-connection is a problem (see \ref{Romani}) only if one wants to interpret the connection as part of the spacetime geometry (for that a $\Spin(\eta)$ connection is needed); but one can also drop the geometric viewpoint for a while and regard the connection as a gauge field.
The $\SU(2)$ connection is then perfectly understandable as a gauge field and in this sense no physical interpretation problem arise.

The problem of the origin of the geometry from the gravitational field is a long standing issue in quantum gravity but the (4d) $\SU(2)$ BI connection establishes a classical scenario that, we believe, is worth discussing.
In fact we have here a gauge field which, together with the frame (and via the extrinsic curvature that is determined as a function of the frame),
uniquely determines a (4d) $\Spin(\eta)$ connection, i.e.~part of the spacetime geometry. As such, the geometry in this framework can be regarded as an emergent structure, already at the classical level. 
Of course, we are not claiming that the problem has been (or it can be, nor it should be) solved at a classical level. One should first understand whether  the emerged spin connection has anything to do with the spacetime geometry which is observationally defined out of the gravitational field; this in turns requires a detailed understanding
of all classical and quantum aspects.
However, we believe it shows a possible mechanism for which the geometry is absent at the quantum level and can emerge at the classical level as a composite structure.

In some sense this would be the strongest implementation of the background independence, showing how quantum gravitational physics can be described 
as a gauge theory where no geometry (not only no background metric) is a fundamental ingredient.


\ms 

The main concern of this paper is therefore to discuss how and under which assumptions the objects can be regarded as global connections of a given group.
Let us mention that locally there is of course no issue: any collection of functions with appropriate indices defines a local connection.
In other words, there are no constraints on local coefficients of a connection; when one decides to work locally in a sigle chart there is no issue to be discussed.

Discussing globality of a connection means instead checking that one gets not only a sheaf of local coefficients for any chart of an atlas covering spacetime, but that this sheaf is compatible with 
the equivalence relation induced by transformation rules. In other words, one should check that the correct transformation rules hold for the local coefficients.
Equivalently, one should define a principal bundle and a global connection on it that locally induces the sheaf of coefficients.

For a $\Spin(\eta)$ connection on $M$ one should thence define coefficients $\om^{ab}_\mu$, antisymmetric in the upper indices $[ab]$, on each patch of an open covering of spacetime. In two overlapping patches the coefficients cannot be chosen independently since in the overlapping they should match as
$$
\om'^{ab}_\mu=\bar J^\nu_\mu \ell ^a_c \( \om^{cd}_\nu \ell^b_d + d_\nu \bar \ell^c_d \eta^{db}\)
\fl{SpinConnectionTR}$$
for some $\Spin(\eta)$--valued pointwise function $\ell^a_b$ defined on the patches overlap, being $J^\nu_\mu$ the Jacobian and the bar denoting inverse matrices; Greek indices run from $0$ to $3$.

Analogously, a $\SU(2)$ connection is defined by local coefficients $A^k_\mu$ which match on the overlaps as follows
$$
A'^k_\mu= \bar J^\nu_\mu \(\la^k_j\>A^{j}_\nu-\frac[1/2]\ep^{kj}{}_{i} \la^i_l\di_\nu \bar\la^l_j\)
\fl{SUTransfromationRules}$$
for some $\SU(2)$--valued pointwise function $\la^i_j$ defined on the patches overlap.
These definitions can be easily extended to global connections on $S\subset M$. 

Let us remark that the functions $\ell^a_b$ (as well as $\la^i_j$) define a $2$-cocycle valued in the relevant group and hence define a (unique up to isomorphisms) principal bundle with $\Spin(\eta)$ (or $\SU(2)$) as structure group on which the global connection is defined.

Now, if one tries to define a spacetime BI connection as
$$
A^k_\mu= \ga \om^{0k}_\mu+ \frac[1/2] \ep^k{}_{ij}\om^{ij}_\mu
\fl{SpacetimeBIConnection}$$
mimicking \ShowLabel{SpatialBIConnection} and assuming that $\om^{ab}_\mu$ transform as expected for the coefficients of a global $\Spin(\eta)$ connection,
one simply does not get the expected transformation rules for $A^k_\mu$.

Actually the same happens also on $S$ for \ShowLabel{SpatialBIConnection}; but in this case topological reasons ensure that the principal bundle on which the connection is defined is trivial and any set of coefficients defines in any case a global connection. However, we cannot be satisfied with this situation; one should in fact  specify
which trivialization has been used in \ShowLabel{SpatialBIConnection} since the resulting connection depends on the trivialization.
In other words, even if a global connection is obtained it should be required to obey transformation laws \ShowLabel{SUTransfromationRules}
with respect to {\it active gauge transformations} and this is not the case.
In any event, without other specific assumptions the BI field is not a connection; this means that the definition of its holonomy (as well as the whole quantization procedure) is strongly questionable: it lacks in fact of rigorous basis and it could eventually fail to be feasible.

In other words one is not trying here to define a $\SU(2)$ connection but rather a map between $\Spin(\eta)$ connections and $\SU(2)$ connections;
accordingly such a map must be compatible with transformation rules and \ShowLabel{SpacetimeBIConnection} is not.
Thence it is necessary to investigate when or how BI field is a connection.

\ms

In Section $2$ we shall precisely show why BI fields do not transform as a connection under active gauge transformations.
We shall also determine additional assumptions for obtaining the correct behaviour.

In Section $3$ we shall introduce reductions of the structure group of a principal bundle which provide the framework in which the additional assumptions of Section $2$ can be implemented. 

In Section $4$ we shall review the dynamics based on Holst's action principle.

In Section $5$ we will provide a brief analysis of Samuel's argument.

\

\NewSection{Barbero-Immirzi Connection}

All we shall say hereafter holds true for both the space and spacetime fields. Accordingly let us omit to indicate the lower coordinate index in order to have one expression for both space and spacetime.

Let us also define the BI field and the {\it BI extrinsic field} by setting 
$$
A^k= \ga \om^{0k}+ \frac[1/2] \ep^k{}_{ij}\om^{ij} 
\qquad\qquad
K^k :=  \om^{0k}
\fl{BIAK}$$

Let us notice that the new fields $(A^k, K^k)$ are as many as the old ones $\om^{ab}$ (both over $M$ and over $S$);
accordingly that is just a new set of field coordinates.
Both fields $(A^k, K^k)$ are defined as a function of $\om^{ab}$ for which transformation rules have been specified; as a consequence one can {\it compute}
transformation rules of  $(A^k, K^k)$ which result uniquely determined.

Let us consider a gauge transformation of the group $\Spin(\eta)$ locally given by
$$
S'= \vp(x)\cdot S
\fn$$
Since the group $\Spin(\eta)$ is a double covering of the relevant orthogonal group $\SO(\eta)$, the covering map $\ell:\Spin(\eta)\arr \SO(\eta)$ defines 
a gauge transformation of the group $\SO(\eta)$ given by $\ell (x)= \ell\circ \vp(x)$. 
Let us split algebra indices $a=0..3$ into time-spatial indices $(0, i=1..3)$.

The transformation rules obtained for the field $(A^k, K^k)$ are:
$$
\eqalign{
A'{}^i =& \Big[
\(\frac[1/2]\ep^i{}_{jk} \ell^j_m  \ell^k_l \ep^{ml}{}_h
+\ga \ell^0_m\ell^i_l\ep^{ml}{}_h\) A^h+\cr
&+\frac[1/2]\ep^i{}_j{}^k \ell^j_0 \di \bar \ell^0_k
+\frac[1/2]\ep^i{}_j{}^k \ell^j_m \di \bar \ell^m_k
+\ga \ell^0_m\di \bar\ell^m_j\eta^{ji}
+\ga \ell^0_0\di \bar\ell^0_j\eta^{ji}+\cr
&+\ep^i{}_{jk} \ell^j_0\ell ^k_m K^m
+\ga\(\ell^0_0 \ell ^i_h
-\ell^0_h\ell^i_0
-\frac[1/2]\ep^i{}_{jk} \ell^j_m\ell ^k_l\ep^{ml}{}_h
\)K^h+\cr
&-\ga^2\ep^{mj}{}_h \ell^0_m\ell^i_j K^h
\Big]\cr
K'{}^i=& \Big[
\(\ell^0_0\ell^i_k-\ell^0_k\ell^i_0\)K^k
-\ga \ell^0_k\ell^i_j\ep^{kj}{}_l K^l
+ \ell^0_k\ell^i_j\ep^{kj}{}_l A^l+\cr
&+\(\ell^0_0\di \bar\ell^0_j +\ell^0_k\di \bar\ell^k_j\)\eta^{ji}
\Big]\cr
}
\fl{BadTransformationRules}$$

We stress that $\ell^0_0$, $\ell^i_0$, $\ell^0_i$, $\ell^i_j$ denote the blocks of $\ell^a_b\in \SO(\eta)$
and hence no specific form can be assumed in general.
One can try with some explicit generic element of $\Spin(\eta)$ to show that extra terms in 
\ShowLabel{BadTransformationRules} do not vanish in general.
Because of this, one cannot assume $A^k$ to be a global $\SU(2)$-connection; in fact one cannot understand it as a separate field with respect to $K^k$:
they are not adapted to eigenspaces of the representation of the group $\Spin(\eta)$ and as such they cannot either be interpreted as two independent fields.

However, let us notice a nice algebraic fact: {\it if we  could} restrict the spin group to
a subgroup $G\subset \Spin(\eta)$ isomorphic to $\SU(2)$ for which elements $\si\in G$ project over the elements of $\SO(\eta)$ in the form
$$
\ell(\si)=\(\matrix{
1 & 0 \cr
0 & \la(S_+)\cr
}\)
\fl{SU2SpecialForm}$$
the transformation rules \ShowLabel{BadTransformationRules} would drastically simplify as we shall see below in detail.

In the Euclidean case we can consider $i:\SU(2)\arr \Spin(4): S_+\mapsto (S_+, S_+)$ 
where we are using the canonical isomorphism $\Spin(4)\simeq \SU(2)\times \SU(2)$.
In the Lorentzian case, $\Spin(3,1)\simeq \SL(2, \C)$ and use the canonical embedding $i:\SU(2)\arr \SL(2, \C)$.
In both cases the image of $\ell \circ i:\SU(2)\arr \SO(\eta)$ is in the form \ShowLabel{SU2SpecialForm}.

For elements in this simpler form the extra terms in \ShowLabel{BadTransformationRules}
do in fact vanish and the transformation rules obtained are the appropriate ones for a $\SU(2)$-connection (see equation \ShowLabel{SUTransfromationRules}) and an (algebra valued) covector:
$$
\eqalign{
A'{}^i =& \Big[
\frac[1/2]\ep^i{}_{jk} \la^j_m  \la^k_l \ep^{mn}{}_h A^h
+\frac[1/2]\ep^i{}_j{}^l \la^j_m \di \bar \la^m_l
+\ga\(\la^i_h
-\frac[1/2]\ep^i{}_{jk} \la^j_m\la^k_l\ep^{ml}{}_h
\)K^h\Big]=\cr
=&\Big[
\la^i_j  A^j
+\frac[1/2]\ep^i{}_j{}^l \la^j_m \di \bar \la^m_l
+\ga\(\la^i_h
-\la^i_h
\)K^h\Big]=\cr
=&\Big[
\la^i_j  A^j
+\frac[1/2]\ep^i{}_j{}^l \la^j_m \di \bar \la^m_l\Big]\cr
K'{}^i=&  \la^i_j  K^j\cr
}
\fl{GoodTransformationRulesBI}$$
Hence we should only investigate when and under which conditions one is allowed to consider the subgroup of gauge transformations in the form $\si(x)\in G
\subset \Spin(\eta)$.

The issue is not trivial since the local expression for a $\Spin(\eta)$-gauge transformation $\phi(x)$ as $\phi(x)=i(S_+(x)) $ does in fact depend on the trivialization chosen on $P$.
Even tuning $\phi(x)=i(S_+(x))$ in a given trivialization this form has no intrinsic meaning; 
when the trivialization is changed  the special form is not preserved in general.

In fact, for example in Euclidean signature, transition functions of $P$ are in general of the form $(\vp_+, \vp_-)$ so that, in the new trivialization, the same gauge transformation is generated by $(\vp_+\cdot S_+, \vp_-\cdot S_+)$ which is no longer  in the special form.

The only case in which the special subgroup is intrinsic is when $P$ has some special trivialization
with transition functions in the special form $i(\vp_+)\in \Spin(\eta)$.
When this happens one says that $P$ admits a {\it reduction} from the group $\Spin(\eta)$ to the group $\SU(2)$, or in short a $\SU(2)$-reduction; see \ref{KobaNu}. 
This corresponds to require that one can cover the whole spacetime with patches choosing a local gauge in each patch such that all transition functions among different local gauges are in the special form $i(\vp_+)\in G\subset \Spin(\eta)$. 

Of course one could {\it assume} $P$ to have such $\SU(2)$-reduction, which usually restricts the allowed $P$ and possibly imposes topological restrictions also on $M$.

\NewSection{Reductions}

Let $P$ be a $\Spin(\eta)$ principal bundle over spacetime $M$ and $\om^{ab}_\mu$ be a spin connection of $P$.
Let us assume that $M$ allows metrics of signature $\eta$ and spin structures (in order to allow global spinors); these conditions imply that the first and second 
Stiefel-Whitney class are vanishing.

A $\SU(2)$-reduction of the bundle $P$ is a $\SU(2)$-principal bundle $^+P$ together with a principal morphism $\io: {}^+P\arr P$ with respect to the group embedding $i:\SU(2)\arr \Spin(\eta)$ defined above. One can describe the reduction by means of the following commutative diagram:
$$
\begindc{\commdiag}[1]
\obj(40,80)[+P]{$^+P$}
\obj(110,80)[P]{$P$}
\obj(40,30)[M1]{$M$}
\obj(110,30)[M2]{$M$}
\obj(20,60)[SU]{$\SU(2)$}
\obj(135,60)[Spin]{$\Spin(\eta)$}
\mor{SU}{Spin}{$i$}[\atleft, \dasharrow] 
\mor{P}{M2}{}
\mor{+P}{M1}{}
\mor{+P}{P}{$\io$}
\mor{M1}{M2}{}[\atleft, \solidline] \mor(40,33)(110,33){}[\atleft, \solidline]
\enddc
\fn$$

We stress that when ever such a reduction exists then the bundle $P$ has {\it by construction} a trivialization with transition functions in the special form $i(\vp_+)$. 
In this case, one can induce such a trivialization on $P$ by using any trivialization of ${}^+P$.
In fact the reduction is equivalent to the existence of a trivialization with reduced transition functions.

On the bundle $P$ we can also globally define the subgroup of gauge transformations in the special form $i(\phi_+)$. We shall denote this subgroup by $\Aut({}^+\<\<P)\subset \Aut(P)$ since it is an isomorphic image of the group of all gauge transformations on ${}^+\<\<P$. 

Now if $\om^{ab}_\mu$ is a connection on $P$ we can set
$$
\cases{
&\A^i_\mu=\ga\om^{0i}_ \mu+ \frac[1/2]\ep^i{}_{jk}\> \om^{jk}_ \mu \cr
&\K^i_\mu= \om^{0i}_ \mu\cr
}
\fn$$
Because of the particular form of the transition functions on $P$, by going through what we said above we can consider transformation rules of $(\A^i_\mu, \K^i_\mu)$ with respect to the transformation rules in $\Aut({}^+\<\<P)\subset \Aut(P)$.
By simply resorting to \ShowLabel{GoodTransformationRulesBI} we easily prove that  $\A^i_\mu$ transforms as a $\SU(2)$-connection.
Analogously we can prove $\K^i_\mu$ to be a $\su(2)$-valued covector.

In dimension four the existence of such a reduction is related to the vanishing of the third Stiefel-Whitney class of $M$ (see \ref{Antonsen} and references quoted therein).
Such class is trivial when both the first and the second Stiefel-Whitney classes are trivial (which can be proved by using Steenrod square operators in cohomology; see \ref{Milnor}).
On  the other hand, the first and second Stiefel-Whitney classes of $M$ are already assumed to be trivial to allow spin structures on $M$.
As a consequence, for $\dim(M)=4$ our hypotheses ensure that there are no further obstructions to existence of the required reductions.

The reduction can be now pulled-back to any embedded space manifold $t:S\arr M$ obtaining
$$
\begindc{\commdiag}[1]
\obj(110,80)[+P2]{$^+P$}
\obj(180,80)[hP3]{$P$}
\obj(70,50)[+Si4]{$^+\Si$}
\obj(140,50)[hSi5]{$\Si$}
\obj(110,30)[M2]{$M$}
\obj(180,30)[M3]{$M$}
\obj(70,0)[S4]{$S$}
\obj(140,0)[S5]{$S$}
\mor{+P2}{M2}{}
\mor{hP3}{M3}{}
\mor{+Si4}{S4}{}
\mor{hSi5}{S5}{}
\mor{+P2}{hP3}{}
\mor{M2}{M3}{}[\atleft, \solidline] \mor(110,33)(180,33){}[\atleft, \solidline]
\mor{S4}{S5}{}[\atleft, \solidline] \mor(70,3)(140,3){}[\atleft, \solidline]
\mor{S4}{M2}{}[\atleft, \injectionarrow]
\mor{S5}{M3}{}[\atleft, \injectionarrow]
\mor{+Si4}{+P2}{}[\atleft, \injectionarrow]
\mor{hSi5}{hP3}{}[\atleft, \injectionarrow]
\mor{+Si4}{hSi5}{}
\enddc
\fn$$

The BI connection $\A^k_\mu$ and the extrinsic curvature $\K^k_\mu$ pull-back to
$$
\cases{
&\A^i_A=\ga\om^{0i}_A+ \frac[1/2]\ep^i{}_{jk}\> \om^{jk}_A \cr
&\K^i_A= \om^{0i}_A\cr
}
\fn$$
which locally are {\it exactly} the standard fields defined in LQG but here with a global meaning as a global connection 
and covector on $^+\Si$.
They transform in fact as expected with respect to the group of gauge transformations $\Aut(^+\Si)\subset \Aut(P)$ (which are intrinsically well-defined), while generic spin transformations 
$\Aut(P)$ mix the two fields as in \ShowLabel{BadTransformationRules}.

In other words, the BI connection is a $\SU(2)$-object, not a $\Spin(\eta)$-object.
Accordingly, one of the best perspectives to look at the BI framework is exactly the need to provide  a $\SU(2)$-formulation of GR already at spacetime level, variously dropping or using the antiselfdual part of the spin group.

\NewSection{Holst's Lagrangian}

Once we recognize that reductions allow to define BI connection and that they exist when global spinors are defined, dynamics is required.
It is well-known that dynamics is described by the Holst Lagrangian (see \ref{Holst})
$$
L_\ga( e, j^1\om) = \frac[1/4\ka]R^{ab} \land e^c\land e^d \ep_{abcd} 
+\frac[1/2\ka\ga]R^{ab} \land e_a \land e_b
\fl{HolstLag}$$
where $R^{ab}$ is the Riemann curvature of $\om^{ab}$.

At spacetime level we defined a change of field coordinates $(e, \om)\mapsto (e, K, A)$ and we can pull-back the Holst Lagrangian \ShowLabel{HolstLag}
to write it down in the new coordinates; see \ref{ourHolst}.
Then in the Hamiltonian framework some field equations define constraints that express the extrinsic curvature $K= \ga^{-1}\(A- \Ga\)$, where $\Ga$ is the connection induced by the frame, and the standard constraints
$$
\cases{
&\na_A E^A_i=0\cr
&F^i_{AB} E_i^A=0\cr
&\ep_k{}^{ij}F^k_{AB} E_i^A E_j^B
-2(\si^2-\ga^2)K^{[i}_A K^{j]}_B E_i^A E_j^B=0\cr
}
\fn$$
where $\si^2=1$ in the Euclidean case and $\si^2=-1$ in the Lorentzian case.
These constitute in fact the starting point of LQG quantization procedure.

\ 

\NewSection{Samuel's Argument}

Samuel presented an argument to show that BI connection is not the space pull-back of a spacetime spin connection; see \ref{Samuel}.
The argument is based on a specific example of computation of the trace of holonomy along a specific path in order to show that it depends on the Immirzi parameter. We have here the necessary tools to revisit the example in view of a detailed analysis of Samuel's example and interpretation of the $\SU(2)$
covariance introduced by the reduction.

Let us consider Minkowski spacetime $M\equiv \R^4$ in spatial spherical coordinates $\eta= -dt^2 + dr^2 + r^2 \( d\te^2 + \sin^2\te d\vp^2\)$.
Besides the standard slides $S_\tau=\{t=\tau\}$, which are related to the standard orthonormal frame $(e_0=\del_0, e_1=\del_r, e_2=r^{-1}\del_\te, e_3= (r\sin\te)^{-1}\del_\vp)$, one can consider a hyperbolic slicing
$$
S'_\tau=\{ t=\tau+ \cosh\rho, r=\sinh\rho)\}
\fn$$
which is related to the frame
$$
\cases{
&e'_0=\sqrt{1+r^2} e_0 + re_1\cr
&e'_1=r e_0 + \sqrt{1+r^2}e_1\cr
}
\qquad\qquad
\cases{
&e'_2=e_2\cr
&e'_3=e_3\cr
}
\fn$$

The two frames are related by a pointwise Lorentz transformation, namely the following block matrix
$$
\ell_a^b= \(\matrix{
\matrix{ \sqrt{1+r^2} & r\cr
r& \sqrt{1+r^2} \cr
}& \Bigg|
&\O \cr
\hbox{---------------------------}&& \hbox{---}\cr
\O&\big|&\one\cr
}\)
\fn$$
We stress that this Lorentz matrix is not in the form \ShowLabel{SU2SpecialForm}, hence it is induced by a matrix in $\SL(2,\C)$ which is not in $\SU(2)\subset\SL(2,\C)$.

One can consider the spin connections induced by the two frames; let us denote them by $\om^{ab}_\mu$ and $\om'^{ab}_\mu$, which are of course related by
\ShowLabel{SpinConnectionTR}. Then one can change field coordinates and define two pairs of fields $(A^k_\mu, K^k_\mu)$ and $(A'^k_\mu, K'^k_\mu)$.
Accordingly, one should not expect the two BI connections $A'$ and $A'$ to be gauge equivalent, as it should be if the Lorentz transformation were in the special form \ShowLabel{SU2SpecialForm}.

Since $A$ and $A'$ are two different $\SU(2)$ connections there is no reason to expect the trace holonomies to be the same (as one should expect if they were spin conections). How and why the two set of fields $(A^k_\mu, K^k_\mu)$ and $(A'^k_\mu, K'^k_\mu)$ represent the same physical situation (as one should conclude in view of the hole argument) is hard to say. Whether there is a fundamental difference between $\SU(2)$ gauge transformations and general $\Spin(\eta)$ transformations needs further investigation. It may seem promising that a generic Lorentz transformation does not preserve the foliation and hence have to do with synchronization protocols, while $\SU(2)$ transformations do in fact preserve the foliation and only change the spatial observer's frame (i.e.~conventions).

\NewSection{Conclusions and Perspectives}

We have provided a global geometric framework to introduce the BI connection and understand its global properties.
We have also shown that the BI connection does in fact appear as the restriction of a global $\SU(2)$-connection defined on the whole spacetime.
The construction does not rely on the possible triviality of the principal bundle which encodes the gauge structure  of the model nor it resorts to gauge fixings which would spoil manifest gauge covariance.
On the contrary, the construction relies on the existence of a $\SU(2)$-reduction which is the correct mathematical structure to be considered.

We believe that this framework might help to investigate the global gauge structure of the theory and the relations among different gauge groups $	\Spin(4)$, $\Spin(1,3)$, $\SU(2)$ which appear in LQG.
These groups encode the covariance properties of GR and a better control on their mutual relations might provide a suitable framework to clarify the covariance issues which are sometimes still under 
discussion in LQG.

Finally, the spacetime interpretations of the objects appearing in LQG might help in clarifying the issues connected to the semiclassical limits of LQG itself.

\Acknowledgements

This work is partially supported by MIUR: PRIN 2005 on {\it Leggi di conservazione e termodinamica in meccanica dei continui e teorie di campo}.  
We also acknowledge the contribution of INFN (Iniziativa Specifica NA12) and the local research funds of Dipartimento di Matematica of Torino University.

\ShowBiblio


\end